\documentclass[runningheads]{llncs}
\usepackage{graphicx}
\usepackage{booktabs}
\usepackage[T1]{fontenc}
\usepackage[utf8]{inputenc}
\usepackage{graphicx}
\usepackage{microtype}
\usepackage{cite}
\usepackage[hyphens]{url}
\usepackage{amssymb}
\usepackage{amsmath}
\usepackage[mathscr]{euscript}
\usepackage{multirow}
\usepackage{tabularx}
\usepackage{diagbox}
\usepackage{adjustbox}
\usepackage{subfigure}
\usepackage{tikz}
\usepackage{enumitem}

\usepackage[misc,geometry]{ifsym}
\usepackage{outlines}

\usetikzlibrary{fadings}
\usetikzlibrary{patterns}
\usetikzlibrary{shadows.blur}
\usetikzlibrary{shapes}
\usepackage[colorlinks = true, allcolors=blue]{hyperref}
\usepackage{orcidlink}

\begin{document}
\title{SocioHub: An Interactive Tool for Cross-Platform Social Media Data Collection}
\titlerunning{SocioHub: A Cross-Platform application}
%
\authorrunning{A. Nirmal et al.}
\author{Ayushi Nirmal\textsuperscript{(\Letter)}\orcidlink{0009-0001-9091-7255} \and
Bohan Jiang\orcidlink{0000-0001-8552-2681} \and
Huan Liu\orcidlink{0000-0002-3264-7904}}

%
%

\institute{Arizona State University, Tempe AZ 85281, USA \\
\email{\{anirmal1, bjiang14, huanliu\}@asu.edu}}

%
\maketitle              
\begin{abstract}
Social media is inherently about connecting and interacting with others. Different social media platforms have unique characteristics and user bases. Moreover, people use different platforms for various social and entertainment purposes. Analyzing cross-platform user behavior can provide insights into the preferences and expectations of users on each platform. By understanding how users behave and interact across platforms, we can build an understanding of content consumption patterns, enhance communication and social interactions, and tailor platform-specific strategies. We can further gather insights into how users navigate and engage with their platforms on different devices. In this work, we develop a tool \texttt{SocioHub}, which enables users to gather data on multiple social media platforms in one place. This tool can help researchers gain insights into different data attributes for users across social media platforms such as Twitter, Instagram, and Mastodon.

\keywords{Social Media Platforms  \and Twitter \and Instagram \and Mastodon.}
\end{abstract}

\section{Introduction}

Cross-platform analysis closes a methodological gap by acknowledging the interaction between the social phenomenon being studied and the medium in which it is being studied. This helps to highlight the various affordances and cultures of online platforms. Cross-platform analysis offers increased chances to consider how social media data are influenced by the structures and affordances of the platforms themselves as well as the social phenomena under consideration~\cite{ref_1}. Researchers should exercise caution when comparing examples of the same social phenomenon across different platforms because of the significance of each platform's unique characteristics and structures.

Cross-platform studies must pay close attention to both ``Medium research'' and ``Social research,'' in particular. Platform affordances like hashtag usage cultures and moderation practices are covered in Medium research, as well as platform architecture like how platforms present information on screens and provide data to researchers through APIs. Through this work, we have tried to bring three platforms together on \texttt{SocioHub} by getting insights into comparable attributes for user information across different social media websites~\cite{ref_2}. With users playing around various social media platforms, their attention is fragmented. It is no longer sufficient to rely solely on data from a single platform to understand user behavior~\cite{ref_5}. Cross-platform analysis provides a more comprehensive view of their online behavior holistically. Protecting user privacy was a primary consideration when creating this tool, therefore, we only considered publicly available profiles.

\section{Methodology}
\begin{figure}
\includegraphics[width=\textwidth]{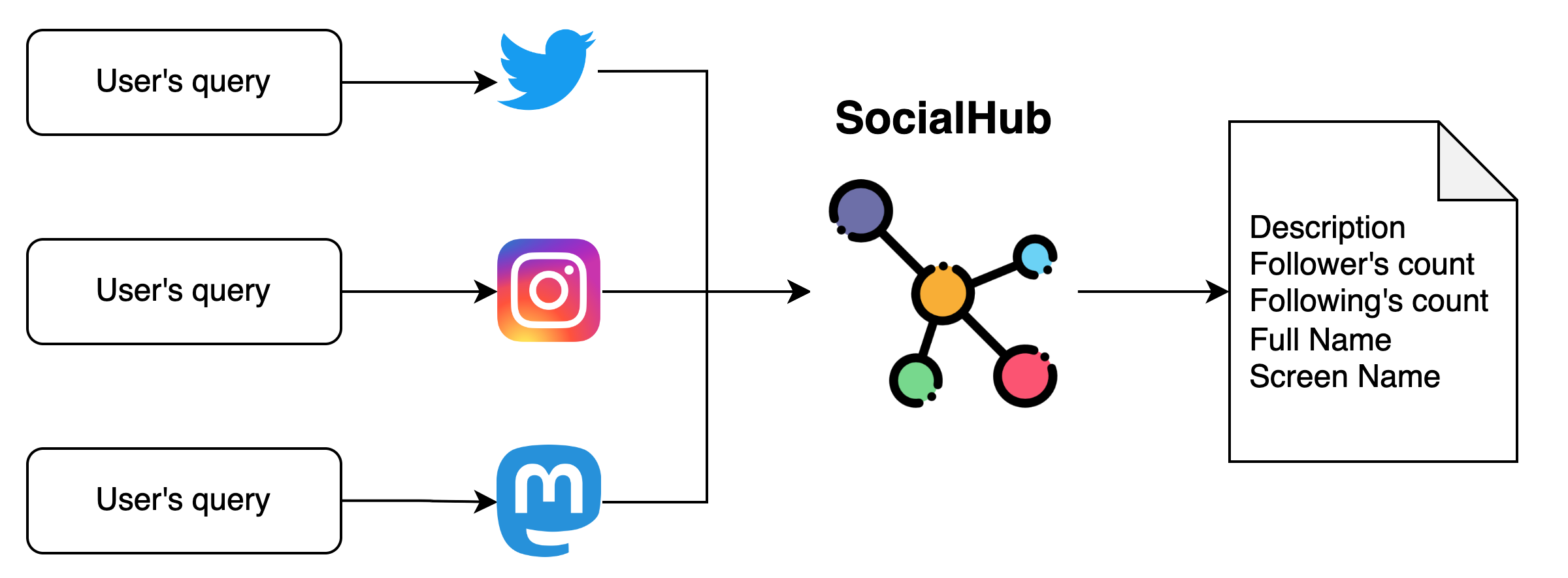}
\caption{Architecture of SocioHub, a cross-platform tool for user interaction.} \label{architecture}
\end{figure}

\begin{figure}[htb]
\includegraphics[width=\textwidth]{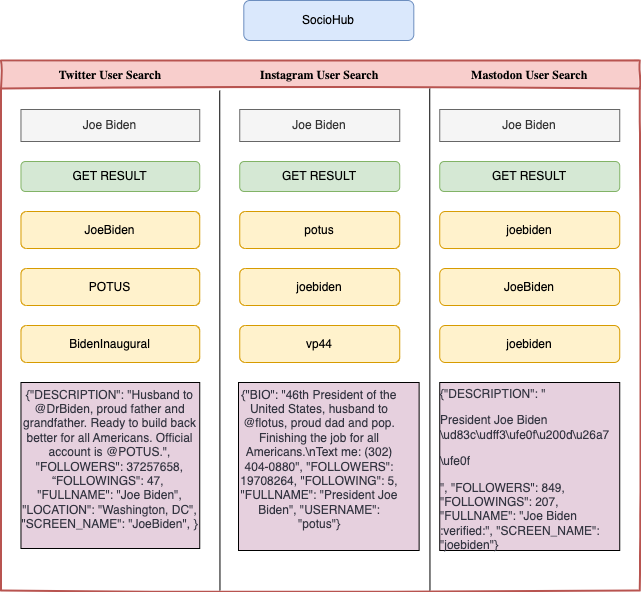}
\caption{Description of users across three platforms.} \label{SocioHub}
\end{figure}

Our aim was to study three social media platforms, including Twitter, Instagram, and Mastodon, in order to identify common data attributes that could be compared across these platforms. To accomplish this, we created a tool called ``SocioHub," shown in Figure \ref{SocioHub}, which enables researchers to easily search for and interact with users on all three platforms as shown in Figure~\ref{architecture}. The primary goal of SocioHub is to allow researchers to examine user behavior in response to the same search query across multiple platforms. By collecting comparable data attributes from each platform, we enable users to conduct detailed analyses of the differences and similarities between these platforms.

\subsection{Architecture}

Once the essential data attributes were identified, we established various methods to enable users to search for individuals based on the closest matching string~\cite{ref_3}. The tool is designed to facilitate user input of a search query for a username, and it subsequently provides a list of the most relevant users based on the platform being utilized. Each query that returns usernames includes comprehensive details about the users, such as their screen name, display name, description, number of followers, and number of followings. 

\begin{table}
\centering
\begin{tabular}{|l|p{3mm}l|}
\toprule
{\large \bfseries Social Media Platform} & & {\large \bfseries Credentials Required}\\
\hline
Twitter & &  API Key, API Key Secret, \\
        & & Access Token, Access Token Secret \\
        \hline
Instagram & & Username, Password \\
\hline
Mastodon & & Access Token, Base URL \\

\bottomrule
\end{tabular} \newline

\caption{Authentication Keys Required for Platforms Used.}\label{tab1}
\end{table}

The primary objective of constructing the tool is to examine the common features and gather insights from data attributes for the queried users. This includes assessing how the number of followers a user has on one platform differs on another social media platform for the same user. Additionally, it involves determining the variation in user visibility based on their connections on different social networking sites. We have taken into account both centralized social networking sites like Twitter~\cite{ref_4,ref_5,ref_6} and Instagram~\cite{ref_7}, as well as decentralized ones like Mastodon~\cite{ref_8}. This comparative analysis serves as a study to understand the patterns of user behavior across different types of social networking sites.

Through various authentication mechanisms that vary based on the sites utilized, various social media networks offer insight into monitoring user activity or content usage. The list of tokens and authentication requirements necessary to develop this tool for doing analysis on the social network sites mentioned are provided in Table \ref{tab1}. These token values can be changed to see a different aspect from a different user.

\subsection{Data Description}

The data attributes for a user vary depending on the social media platform. However, for our analysis, we focused on comparable fields across all the social media platforms we considered as mentioned in Table \ref{tab2}. One important data attribute we included in the comprehensive detailed section for Twitter user results is location, as we considered it important for the study. Therefore, in general, we retrieve a total of “five” data attributes about a user to provide a brief overview of their profile. The only exception is Twitter, where we also include the location attribute, recognizing its importance. Since we only took into account publicly visible profiles, protecting user privacy was a top priority when developing this tool.

\begin{table}
\centering
\begin{tabular}{|l|p{3mm}l|}
\toprule
{\large \bfseries Social Media Platform} & & {\large \bfseries Data Attributes Collected}\\
\hline
Twitter &  & name, screen\_name, description, followers\_count, \\
        & & friends\_count, location \\
\hline
Instagram & & full\_name, username, biography, followers, \\
        & & followees \\
\hline
Mastodon & & display\_name, username, note, followers\_count, \\ 
        & & following\_count \\
\bottomrule
\end{tabular} \newline

\caption{Data Attributes for Social Media Platforms Used.}\label{tab2}
\end{table}

We also store the query and its associated results in the MongoDB database after successfully retrieving the user information from all three platforms. The query search vs results can be exported to see the records at a later time.


\section {Conclusion and Future Work}

Through our study, we have attempted to show how combining several social media platforms can provide us with a better knowledge of user behavior on various platforms. Our work is simply a prototype of how diverse platforms might be merged and researchers can receive insights from a single data source. Depending on the applications chosen for study or analysis to collect data from social media platforms, it can also be expanded to other various platforms. We encourage the development of a broad tool that can expand this functionality and further incorporate more data attributes, which could add to the importance of the research work on the analysis of cross-platform social media platforms, despite the limitations to our work caused by the use of limited rate limits on platforms provided APIs.

Furthermore, we strongly recommend that user policy must be followed while adhering to the company’s policies when trying to analyze social platforms. We also want to draw your attention to the fact that this work was created prior to Twitter enacting its policies about changes to API usage and charges. It is crucial to protect individuals' privacy and abide by the law and ethical standards when handling personal data. Prioritizing privacy and managing personal information in an ethical and legal way should always be priorities. We hope our work can provide valuable insights to future researchers in related fields.

\section*{Acknowledgement}
This work is supported by the Office of Naval Research under Award No. N00014-21-1-4002.

\end{document}